# Understanding of complex periodic transformations of moving domain walls in magnetic nanostripes

JUN-YOUNG LEE, KI-SUK LEE, SANGKOOK CHOI, KONSTANTIN Y. GUSLIENKO, and SANG-KOOG KIM*

Research Center for Spin Dynamics & Spin-Wave Devices and Nanospintronics Laboratory, Department of Materials Science and Engineering, Seoul National University, Seoul 151-744, South Korea
*e-mail:sangkoog@snu.ac.kr

The magnetic field (or electric current) driven domain-wall motion in magnetic nanostripes is of considerable interest because it is essential to the performance of information storage and logic devices. One of the currently key problems is to unveil the complex behaviors of oscillatory domain-wall motions under applied magnetic fields stronger than the so-called Walker field, beyond which the velocity of domain walls markedly drops. Here, we provide not only considerably better understandings but also new details of the complex domain-wall motions. In a certain range just above the Walker field, the motions are not chaotic but rather periodic with different unique periodicities of dynamic transformations of a moving domain wall between the different types of its internal structure. Three unique periodicities found, which consist of different types of domain wall that are transformed from type one to another. The transformation periods vary with the field strength and the nanostripe width. This novel phenomenon can be described by the dynamic motion of a limited number of magnetic topological solitons such as vortex and antivortex in nanostripes.

Magnetic domain walls (DWs) of various internal structures are typically observed in magnetic materials[1-3]. The DWs are ten- or hundred-nanometer-size transition regions in which the orientation of local magnetizations **M** gradually changes between neighboring magnetic domains of different **M** orientations in an unsaturated magnetization state[1,2]. The different structures of DWs vary with material physical parameters and geometry. In particular, in magnetic films and, especially, in patterned magnetic films such as stripes, dots, etc. more complicated DW structures appear since magnetostatic interaction is dominant in the restricted geometry. In the case of magnetic nanostripes, head-to-head transverse- or vortex-type DWs are present, as calculated[4] and observed by various experimental techniques[5-7]. Similar to quantized spin wave eigenfrequencies in patterned films[8] it is expected that moving DWs in nanostripes could exhibit well defined oscillating behaviors due to geometrical confinements.

When an external static magnetic field **H** is applied to a single DW present in magnetic nanostripes, the DW propagates like an independent entity (as a classical Newtonian particle) to reduce the Zeeman energy[5,6,9,11]. In principle, the stronger **H**, the faster the DW motion, and hence the DW velocity $\bar{v}$ increases linearly with increasing **H**. This is true, however, only for cases of relatively low **H**. In stronger fields than a certain threshold field known as the Walker field $H_w$, the velocity does not increase but rather rapidly decreases[9-14]. Such a remarkable drop of the average velocity $\bar{v}$ of DWs moving in nanostripes was surprisingly unexpected, but recently is known to be associated with the oscillatory turbulent motion of DWs[11,14]. Although this phenomenon has been qualitatively explained by one-dimensional (1D) DW model[12] and some dynamic changes of the internal DW structures between the transverse wall (TW) and antivortex wall (AVW) types in the turbulent regime have been numerically investigated[11,14], the underlying physics of the complex DW motion has not been unveiled yet in two-dimensional (2D) cases such as nanostripes of submicron-size width. Thus, understanding the DW motions in a practically applicable 2D nanostripe patterned thin film system is currently a key challenging problem to be solved[10]. This problem is of special importance in the light of the recent achievements of experimental investigations of spin-polarized current driven DW motion[15-18]. The oscillatory DW motion can be excited by current pulses of sufficient strength[15,18]. Well defined nanosecond-oscillations of the nanostripe resistance were observed, which are presumably associated with the periodic change of DW structures[18], but need an adequate physical description.

Here we report an unprecedented finding of various-type oscillatory motions associated with dynamically periodic transformations between different types of moving DWs with the characteristic unique periodicities in the turbulent regime, as studied by micromagnetic calculations on Permalloy (Py) nanostripes of various widths (see Methods). The unique periodicities vary with the width of the nanostripes and the strength of the static applied **H** above $H_w$. The results provide a markedly better understanding of the complex dynamic motions of DWs in the turbulent regime in terms of the emission and absorption of a limited number of moving magnetic solitons with integer and fractional topological charges at both edges of the nanostripes. We formulate the problem of the high-speed dynamics of DWs in terms of motion of magnetic solitons. We answer the questions how and why such dynamically periodic transformations occur in the intermediate fields above $H_w$.

In our simulations, we used each type of the equilibrium head-to-head DW, which were positioned at the center of long nanostripes of different width $w$ for the same length $l$=6.0 μm and thickness $h$=10 nm, as shown in Fig. 1a (see Methods). The local **M** configuration of the internal DW structure at equilibrium with $H$= 0 varies with $w$, that is, TW type for $w < w_c$ whereas vortex wall (VW) type for $w > w_c$, where $w_c$ =152 nm for the given nanostripe height $h$=10 nm[4,13]. Figure 1 shows, for example, just the two cases of $w$=140 and 240 nm, which were selected from all of the cases we considered in the simulations. Upon the application of $H$ along the long axis of the nanostripes, that is, in the +$x$ direction as noted in Fig. 1a, each type of the initial DW starts to move toward the **H** direction, because the area of local **M**, being parallel to the field direction, becomes larger in order to reduce the Zeeman energy. As a result, the movement of the DWs along the nanostripes can be considered as the movement of a classically moving independent object.

The $\bar{v}$ values of DWs moving along the nanostripes were plotted versus the $H$ for the two cases of $w$=140 and 240 nm in Fig. 1b. The motions of the equilibrium DWs in their initial states are relatively simple at small $H$ values. The linear increase of $\bar{v}$ with increasing $H$ up to $H_w$ can be described quite well by a "rigid" DW model[12]. However, just above $H_w$ (here approximately $H_w \sim$ 10 Oe), the $\bar{v}$ decreases remarkably in response to the higher fields, as is known from the literature[11,12]. The two regions are clearly distinguished by the value of $H_w$ below and above which the linear and almost constant response of $\bar{v}$ with $H$ are evident. Also, there is an additional region where $\bar{v}$ increases again with further increase of $H$. Those characteristic DW motions in the three different regions (indicated by I, II, and III) can be briefly described based on the representative snapshot images of the normalized perpendicular magnetization components $M_z/M_s$ ($M_s=|M_s|$) along with the in-plane **M** orientations for the case of $w$=140 nm (insets of Fig. 1b) as well as the DW displacement-versus-time curves (hereafter denoted as D-vs-t curves) for both cases of $w$=140 and 240 nm (Fig. 1c).

In Region I, a single DW (TW for $w$=140 nm and VW for $w$=240 nm) moves steadily along the entire length of the nanostripe (see the first column in Fig. 1c). The $\bar{v}$ of the VW type ($\bar{v}$=80 m/s) is much slower, by several times, than that of the TW one ($\bar{v}$=240 m/s). In Region II, the oscillatory D-vs-t curves represent the periodic oscillatory motions of DWs. The $\bar{v}$ values can be negative in a different periodic manner for each case, indicating that the DWs move backward in the direction opposite to the initial propagation direction (**H** direction). In addition, the time period of the oscillatory motion becomes shorter with further increasing $H$. More interestingly, the shapes of the individual curves in each periodicity differ greatly with changing $H$ and $w$ (see Fig 1c). This phenomenon is related to the oscillatory transformations of DWs of different type and sequential order in response to static external field. In Region III, the motions of DWs are neither steady nor periodic, but rather chaotic with the appearance of multiple vortex (V) and antivortex (AV) states (Fig. S3b) during the magnetization reversal, which leads too much faster $\bar{v}$ than in Region II[19].

Such oscillatory DW dynamics occurring above $H_w$ has been explained within the framework of the 1D model of bulk materials[12], where the 1D Bloch wall periodically changes its polarization (the average magnetization in the transverse direction) via the transformation to the Néel wall and back[20]. Also, the numerical calculations of DW dynamics in 2D nanowires[18,21,22] show that the DW motion is also periodic above $H_w$, but that the internal states in the course of DW oscillations are absolutely different from the prediction of Ref. 12. Meanwhile, recent studies have also demonstrated the complexities of the DW dynamic behaviors[6,7,11,18,22,23], such as the occurrence of negative differential mobility (defined as $\mu = v/H$) above $H_w$, large contrasts in mobility in low and high magnetic fields, and different $H_w$ values depending on the thickness, width, magnetic anisotropy, damping, and edge roughness of nanowires[8,11,23]. However, our simulation results, more interestingly, reveal that not only the shape of the oscillatory D-vs-t curves depends on $H$ and $w$, but also that there exist only three different unique periodicities consisting of the different characteristic shapes of parts of the D-vs-t curves, including the linear, convex-up, and convex-down curves. The combination of these different shapes determines a variety of the overall D-vs-t curves.

To elucidate the underlying physics of how such different-type oscillatory curves occur, we plot in Fig 2a the three unique periodicities of the overall shapes of the various D-vs-t curves along with the variations of the exchange $E_{ex}$, dipolar $E_{dip}$, Zeeman $E_{zeem}$ energies of the DW dynamic states, and their sums ($E_{tot}$ and $E_{dip} + E_{zeem}$) (Fig 2b). Also, snapshot images of the detailed instantaneous DW magnetization distributions of TW, AVW, and VW types are shown in Fig. 2c together with the trajectories of the magnetic soliton core motions in periodic oscillatory transformation regime of the corresponding shapes are indicated by the gray colors in Fig. 2a. In this article, we chose just the three different sets of ($w$, $H$) = (140, 25), (240, 20), and (240 nm, 30 Oe) to show the characteristic unique periods we found from all the oscillatory shapes of the D-vs-t curves. The distinct periodicities are here expressed by Type i), ii), and iii) for the sake of convenience. As revealed by the detailed in-plane **M** dynamic configuration of each type of DW at the given times in the individual regions of the corresponding periodicities, the periodic transformations between the dynamic internal DW structures of the TW, AVW, and VW types obviously occur in response to static field. The motions of the individual DW types are also represented by the trajectories of their core motions. For Type i), the initially stable TW with the V-shaped **M** configuration (noted as $TW_V$ for its polarization) is transformed first to $AVW_{up}$, second to $TW_\Lambda$, then to $AVW_{down}$, and again to $TW_V$ itself. The subscripts in each DW type indicate the corresponding polarizations (V- or Λ-shaped **M** configuration for TW, and up- or down-core orientation for either VW or AVW). For Type ii), the $VW_{down}$ in the initial state ($H$= 0) is transformed to $TW_V$, $VW_{up}$, $TW_\Lambda$, and again to $VW_{down}$ itself in this sequential order, and hence the dynamic transformation of $TW_V$, $VW_{up}$, $TW_\Lambda$, $VW_{down}$ is repeated in this structural change and sequential order as one period. In contrast to these two types, for Type iii), the initial VW$_{down}$ is transformed to TW$_V$, AVW$_{up}$, TW$_Λ$, and to VW$_{down}$. Quite interestingly, half of each period of Type i) and ii) appears alternately in one period, such as in ①, ② of i) and ③, ④ of ii) in this sequence, as seen from Fig 2c. The AVWs or VWs switch their polarizations of either the up- or down-core orientation via their dynamic transformation into the TW type with either polarization. Note that only Type i) was numerically calculated in Ref. 11 without detailed analysis, whereas the Types ii) and iii) are newly found in the present work.

Also, the trajectories of the moving cores of the TW, AVW, and VW types in their individual transformation regions clearly exhibit their characteristic motions: the straightforward linear motion of forward and then backward motion, and vice versa, respectively. These different motions of the TW, AVW, and VW types result in the relevant shapes of the sharp increase, convex-up, and convex-down curves, respectively, in the oscillatory D-vs-t curves. The AVW and VW types contain a single AV and a single V, respectively, inside the corresponding internal DW structures. The AV and V states bear topological charges (vorticities)[24] and have non-zero gyrovectors[25] (see Methods). Namely, the non-zero gyrovectors result in their non-linear gyrotropic (translation) motion in potential profiles influenced individually by $E_{ex}$, $E_{dip}$, and $E_{zeem}$ terms. The sense of the non-linear gyrotropic motion of V and AV in finite-size magnets is determined by the sign of the product of the core polarization $p$ and the soliton topological charge $q$, where $p = +1$ (-1) for the upward (downward) core orientation, and $q = +1$ (-1) for V (AV)[24,26]. Consequently, any type of V and AV follows the senses of counter-clockwise rotation for $qp = +1$ and clockwise rotation for $qp = -1$ in their gyrotropic motions in a potential well[26,27]. Therefore, the V in the VW type moves gyrotropically following its core polarization, as shown in Fig. 2c. In contrast, the AV in the AVW type is of a topological charge ($q = -1$) opposite to that of VWs ($q = +1$), such that for the same core polarization and the same kind of the potential well, the gyrotropic motion of AVs should occur in the direction opposite to that of Vs. But in restricted geometry (nanostripes), the magnetostatic energy plays an important role in forming the potential well for V or potential hill for AV with respect to the middle of the nanostripe (see Fig. 2b). For the nanostripes, the sense of the gyromotion direction is thus determined by the sign of the stiffness coefficient $\kappa = |\kappa|\text{sign}(q)$ in the corresponding potential profile as well as the product of $pq$ (see Methods). Therefore, the direction of the AV or V rotation depends only on the soliton polarization $p$. The AV (V) in the AVW (VW) type rotates gyrotropically counter-clockwise for $p = +1$ or clockwise for $p = -1$, in agreement with the numerical results shown in Fig. 2c.

Related to the corresponding VW and AVW gyrotropic motions with either up- or down-core orientation, there are obvious correlations of the dynamic transformations between the TW and VW (or AVW) types. The V ($\Lambda$)-shaped **M** configuration of the internal structure of TWs always leads to the upward- (downward-) core orientation of VWs or AVWs, to be transformed from the TWs themselves, due to the rotation sense of the V and AV gyrotropic motions, which is determined by $p$ only. Thus, the VW (or AVW) switches its up- and down-core polarization alternatively through the transformation to the TW$_V$ and TW$_\Lambda$ type alternatively as well.

The nucleation sites at either stripe edges, where the cores of TWs with V- or $\Lambda$-shaped **M** configuration are located, also determine whether the VW or AVW type will be created in the next transformation process. The in-plane **M** configuration of the TWs resembles an isosceles triangle with three apexes (see Figs. 3). For the V ($\Lambda$)-shaped polarization, a single apex is located at the bottom (top) edge, whereas the double apexes at the other side are located at the top (bottom) edge. When the single apexes act as the nucleation sites of the cores to be created in the next transformation, the AVW type is always formed. When one of the other double apexes at the same edge side act as the nucleation sites for the cores, the VW type is always created. These correlations are associated with the rotation sense of V and AV gyromotions and the given polarization state of the TW type that is ready to be transformed to the VW or AVW type. The V and AV gyromotions are relatively slow and lead to negative DW velocities due to their characteristic backward motions against the field direction. This behavior, in turn, leads to essential decrease of the average DW velocity.

To understand why such periodic transformations occur in the dynamic DW movements, we consider the variations of differently contributing energy terms based on the detailed **M** configurations of DW types that are transformed from type one to another as well as the trajectories of their cores. For the case of the steady motion of a TW, $E_{dip}$ and $E_{ex}$ do not much change (decrease very slowly), but $E_{zeem}$ decreases significantly during the straightforward movement along the nanostripe while holding the core of the $TW_V$ along the bottom edge in response to **H**. For moving TWs appearing in the dynamically periodic transformation, $E_{ex}$ increases markedly, whereas $E_{zeem}$ decreases markedly due to their straightforward motions. This occurs because the cores of the AVWs or the VWs continuously move inward (emission of AV or V) the nanostripe according to the nucleation process near the nanostripe edges. Once the cores of the AVWs and VWs are well formed at a distance (~ the core size) far away from the edges, they move inside the nanostripes through their characteristic gyrotropic motions, as mentioned above, in which $E_{ex}$ does not much change, but $E_{zeem}$ and $E_{dip}$ change according to the core positions. As a consequence, $E_{zeem}$ for the AVW (VW) type first decreases (increases) and then increases (decreases), because the AVWs (VWs) move first forward (backward) and then backward (forward), through their characteristic gyrotropic motions, as evidenced by their trajectories shown in Fig. 2c. Comparing $E_{dip}$ of the AVW and VW types, we can state that $E_{dip}$ for VW (AVW) reaches its minimum (maximum) when the V (AV) is located in the middle of the nanostripe. Thus, it shows a deep potential well for the VW type and a shallow potential hill for the AVW type. From these energy variations, it can be understood that the increase of $E_{ex}$ is necessary for the nucleation of the cores of VW or AVW inside the nanostripes because the cores bear finite topological charges. To overcome the energy barrier related to this additional $E_{ex}$ in further motion, some excess energy should be provided to the nanostripe. Thus, in lower fields, TWs continue moving toward the field direction to reduce $E_{zeem}$, because the cores of AVW or VW cannot be nucleated inside the nanostripe if the energy barrier is not overcome. But in higher fields the cores of AVW or VW can be nucleated near the nanostripe edges by pushing AV (V) inward the nanostripe. Once the cores of the VWs or AVWs are well formed inside the nanostripes, they continue to move inside it through the V and AV gyrotropic motions. Note that during the gyrotropic motion of AV or V after their nucleation, the total energy $E_{tot}$ and $E_{ex}$ do not much change, that is, the sum $E_{zeem} + E_{dip}$ is almost constant. We see that the single or double apex nucleation site of the V- or Λ-shaped TWs determine the emitted soliton charge $q$, whereas the energy conservation (the shape of the magnetostatic well (hill)) determines the soliton core polarization $p$ and the direction of further gyrotropic motion.

The drastic dynamic changes of the internal DW structures of different types, described above, are non-trivial. Such periodic DW transformations occurring above $H_w$ can be described in terms of the nucleation and annihilation of V or AV at either edge of nanostripes of a certain narrow width, that is, the emission and absorption of topological magnetic solitions. The Vs and AVs have such topological charges as $q = +1$ and $q = -1$, respectively, as mentioned before. In contrast, the TWs have non-zero average transverse magnetizations along the $Oy$ axis, directed "upward" ($TW_V$) or "downward" (TW$_\Lambda$), which was already represented by the V- or Λ-shaped **M** configuration in the $x$-$y$ plane. These configurations lead to half-integer ($q = \pm 1/2$) topological solitons located on the stripe edges (see Fig. 3)[28]. The AVWs and VWs have, excepting the charges related to AV and V states, respectively, also some half-integer charges due to their edge singularities similar to the ones in the TWs, as indicated in Fig 3. The total topological charge of all of the solitons (indexed by $j$) inside the nanostripe does not change, that is, $\sum_j q_j = const$ in the process of the dynamic transformation of each type of DWs. The single apex of the V-shaped TW configuration corresponds to a half-integer topological charge $q = -1/2$ (only half of the soliton core is inside the nanostripe). Through such soliton, only the AV can be nucleated with $q = -1$, changing the edge soliton topological charge from $-1/2$ to $+1/2$. The double-apex nucleation site corresponds to the edge soliton with $q = +1/2$, and only the vortex with $q = +1$ can be nucleated, changing the edge soliton topological charge from $+1/2$ to $-1/2$ (see Fig. 3).

To predict the trajectories and traveling time periods of the motion of individual solitons, we analytically solved the Landau-Lifshitz equation of motion accounting the potential energy of a soliton in the nanostripe (see Methods for details). The trajectories of the V and AV core positions **X**= ($X$, $Y$) in the nanostripe are given as $X = (\kappa/2\lambda H)Y^2 - (\mu/\lambda)Y$ with $\kappa = |\kappa|sign(q)$. The definitions and physical meanings of the parameters $\kappa$, $\lambda$, and $\mu$ are given in Methods. The traveling time of the V (AV) from one stripe edge to the opposite edge can be calculated as $t_w = \pi/\gamma H$, surprisingly, which does not depend in the main approximation on the nanostripe dimensions and intrinsic magnetic parameters. The parameter $\gamma$ stands for the gyromagnetic ratio. The traveling time of the edge solitons (TWs) between the occasions of the emission and absorption of V and AV can be neglected compared to the $t_w$ values of the Vs and AVs. Therefore, the estimated value of the period of the dynamic DW transformations is approximately $T_w^{2D} = 2t_w = 2\pi/\gamma H$, which corresponds to the AV (V) gyrotropic motion from one nanostripe edge and back. This estimated period $T_w^{2D}$ is closely related to the Larmor frequency, $\omega = \gamma H$. We note that this period is essentially shorter than the Walker period obtained from the 1D model[12] $T_w^{1D} = 2\pi(1+\alpha^2)/\gamma\sqrt{H^2 - H_w^2}$, especially just above the threshold field $H_w$. This means that the different-type DW oscillations for the 2D case and restricted geometry are considerably faster than ones for the 1D case in bulk materials.

The analytically estimated dependence of $t_w$ values versus $H$ for the VW and AVW types are compared with those obtained from the simulations for the cases of $w$=140 and 240 nm, as seen in Figs. 4a and 4b, respectively. They are in good agreement, comparing the simulation results fitted to the $1/H$ dependence, $t_w = \pi/\gamma H$. In the periodic transformations of AVWs (or VWs) to the TW type, we can also estimate the traveling time for TWs only during their appearance. As $H$ increases, the traveling time of the TWs decreases as $1/H$ and finally approaches the zero value at $H = \sim 55$ (50) Oe for $w$=140 (240) nm. This implies that the TW type does not appear beyond such onset fields. Note that the case does not occur in higher fields, because the chaotic behavior showing multi-DW states takes place well before the disappearance of the TWs occurs in the periodic DW transformations. The simulation results for the trajectories of the cores of the AVWs and VWs are also fitted to the analytically derived parabolic equation of $X = (\kappa/2\lambda H)Y^2 - (\mu/\lambda)Y$. For the AVW case $\kappa$ is of a negative value whereas for the VW case $\kappa$ is of a positive value, in accordance with the definition of $\kappa = |\kappa|\text{sign}(q)$. The absolute values of the stiffness coefficients $\kappa$ are, in general, different for the AVWs and VWs (see Fig. 4).

Recent experimental observations of DW motion in the nanostripes can be explained by our present calculations. The resistance oscillations due to variable DW position and DW internal structure transformations were measured by Hayashi et al[18]. The eigenfrequencies of these oscillations vary within the range 0.1 ~ 0.5 GHz for the driving fields of 18 ~ 90 Oe above the Walker field ~14 Oe. The DW of $TW_V$ and $TW_\Lambda$ types have the same resistance, and therefore only half-periods (double eigenfrequencies) of the DW transformations involving the Type i), ii) or iii) periodicities can be measured applying this technique[18]. The half of the slope $\Delta f / \Delta H$ of the experimental dependences of DW oscillation frequency vs. field in Fig. 4c and 5c of Ref. 18 is equal to 2.7 MHz/Oe and 2.8 MHz/Oe, respectively. This value is in good agreement with the slope $\gamma / 2\pi$ =2.9 MHz/Oe for Permalloy[2] according to the equation $1/T^{2D} = (\gamma/2\pi)H$.

In general, the creation of the VWs or AVWs in the nanostripes leads to the gyrotropic motion of the V(AV) cores, resulting in their backward and forward motion and the average DW speed slows down. The fundamental understanding of the dynamics of the DW motions along magnetic nanostripes is crucially important in achieving ultimate ultrafast speeds of recording in information storage and of computing in logic devices[29-31]. In order to suppress the velocity breakdown above $H_w$ or even to increase the velocity further, the nucleation of the VWs or AVWs during their dynamic changes from the TW type should be prevented. In this case, a single high-speed TW can propagate along the nanostripe without any transformation to the VW or AVW type.

In conclusion, the micromagnetic simulations of the field driven DW dynamics in magnetic nanostripes of various widths, and the analytical interpretation of the observed periodic transformations of different-type DWs, offer a new understanding of the complex behavior of oscillatory DW motion. The various types of the oscillatory motion are determined by the types of dynamic changes of the internal structures of different DWs, which have the three different unique periodicities. The overall dynamic motion proceeds through the nucleation, gyrotropic propagation, and annihilation of magnetic Vs and AVs in different unique periodic manners. This behavior can be described in terms of the emission and absorption of topological solitons by the edge solitons above a threshold field, by following the conservation of the total topological charge in the nanostripes. Accordingly, the characteristic motions of DWs in the turbulent regime are thus caused not only by the periodic transformations of internal DW structures from the TW type to the AVW or VW type, and back, but also by the backward and forward motion for the VWs or vice versa for the AVWs due to their gyrotropic motion, which gives rise to a significant reduction in the DW average velocities. The obtained results offer not only a new physical understanding of the complex DW dynamics in patterned thin films, but also a theoretical basis to suppress the remarkable drop in DW velocity in fields above the Walker threshold field. The results serve as a new non-trivial example of strictly periodic dynamical response of non-linear system to a steady external perturbation.

METHODS

**Micromagnetic simulations**

Micromagnetic simulations were carried out by procedures similar to those in Refs. 32, 33, assuming rectangular-shaped Permalloy ($Ni_{80}Fe_{20}$; Py) nanostripes of 10 nm thickness and 6 μm total length. Different $w$ values ranging from 60 to 240 nm in increments of 20 nm under $H$ = 5, 10, 15, 20, and 25 Oe across $H_w$ were used (for the entire data, see Fig S1). The material parameters corresponding to Py were as follows: the saturation magnetization $M_S$ = 8.6 × 10$^5$ A/m, the exchange constant $A$ = 1.3 × 10$^{-11}$ J/m, with zero magnetocrystalline anisotropy. The unit cell dimensions of 5 nm × 5 nm × 10 nm with a constant saturation magnetization for each cell, and the Gilbert damping constant $\alpha$ = 0.01 were used in all of the simulations. The OOMMF code was used to numerically calculate the dynamics of the **M** of the individual cells as well as their interactions based on the Landau-Lifshitz-Gilbert equation of motion[33]. To numerically calculate the dynamic motions of DWs in a nanostripe, first we obtained the equilibrium **M** configuration of a head-to-head TW or VW type placed at the middle position of the long axis of the nanostripes. These initial **M** configurations were obtained with the arbitrary configurations of the TW- or VW-like structures at $H$ = 0 available at the given widths, according to the simulation results reported in Ref. 13. Our results for the nanostripe thickness of 10 nm revealed that the static TW (VW) type is stable at $H$= 0 below (above) $w_c \approx$ 152 nm in accordance with Ref. 13, where $w_c$ is the critical nanostripe width. The static TW- or VW-type **M** configurations of the given nanostripe width were then driven to move along the nanostripe by a magnetic field applied along the long axis of the nanostripe in the +$x$ direction.

**Analytical derivation**

The dynamics of the **M** distributions for the motion of magnetic solitons can be described by the ansatz[19] $\vec{M}(\vec{r},t) = \vec{M}(\vec{r} - \vec{X}(t))$, where **X**=(X, Y) is the position of a soliton in the x-y plane. The **M** dependence on the coordinate along the nanostripe thickness is neglected because the nanostripe is thin (10 nm). In our case the V or AV core centers are considered as topological soliton positions. The Landau-Lifshitz equation of motion can be written in the form of the motion of $\vec{X}(t)$, as $\vec{G} \times \dot{\vec{X}} = \partial W / \partial \vec{X}$, where $\vec{G} = G\hat{z}$ is the gyrovector with the gyroconstant, $G = 2\pi M_s hpq/\gamma$. $W(\mathbf{X})$ is the potential energy of a soliton in the nanostripe, which can be written as $W(\mathbf{X}) = \kappa Y^2/2 - \lambda \vec{X} \cdot \vec{H} + \mu \hat{\vec{z}} \cdot \vec{X} \times \vec{H}$. The first term describes the energy variation in the transverse direction of the nanostripe, where $\kappa$ is the stiffness coefficient, the second term corresponds to the motion of the DW as a whole, and the third term describes the motion of the vortex (antivortex) along the direction perpendicular to the field direction. The coefficient $\lambda$ can be directly calculated such that $\lambda = 2M_s wl$. The coefficient $\mu$ is given by $\mu \approx CM_s wl$ for V or $\mu \approx 0$ for AV, where $C$ is the vortex chirality, +1 or -1. For definitions of the parameters $C$ and $q$, see the papers by Guslienko et al[25,26]. We note that the topological charge can be obtained by integration of the V (AV) magnetization distribution in the vicinity of the V (AV) core where $M_z \neq 0$. The half-integer charges $q = \pm 1/2$ mean that only half of the topological soliton (the V or AV core) is located within the nanostripe[28]. By solving the two basic equations of motion for $X$ and $Y$, the trajectory of V or AV in the nanostripe can be represented as $X = (\kappa/2\lambda H)Y^2 - (\mu/\lambda)Y$ with $\kappa = |\kappa| sign(q)$. The traveling time of the V (AV) from one stripe edge to the opposite edge can be obtained from the solution of the equation of motion for $X(t)$ and is $t_w = \pi/\gamma H$. This time does not depend in the main approximation on the nanostripe sizes of $w$, $l$ and the material parameters of $M_s$, $\alpha$.

References

1. Kittel, C. Physical theory of ferromagnetic domains. *Rev. Mod. Phys*. **21**, 541 (1949).

2. Hubert, A. & Schafer, R. *Magnetic Domains* (Springer, Berlin, 2000).

3. Kim, S.-K., Kortright, J. B. & Shin, S.-C. Vector magnetization imaging in ferromagnetic thin films using soft x-rays. *Appl. Phys. Lett*. **78**, 2742-2744 (2001).

4. McMichael, R.D. & Donahue, M.J., Head to head domain wall structures in thin magnetic strips. *IEEE Trans. Magn*. **33**, 4167-4169 (1997).

5. Ono, T. et al. Propagation of a magnetic domain wall in submicrometer magnetic wire. *Science* **284**, 468-470 (1999).

6. Atkinson, D. *et al*. Magnetic domain-wall dynamics in a submicrometre ferromagnetic structure. *Nature Mater.* **2**. 85-87 (2003).

7. Thomas, L. *et al*. Observation of injection and pinning of domain walls in magnetic nanowires using photoemission electron microscopy. *Appl. Phys. Lett*. **87**, 262501 (2005).

8. Bayer, C. et al., Spin Wave Excitations in Finite Rectangular Elements, in *"Spin Dynamics in Confined Magnetic Structures III"* (Springer-Verlag, Berlin, 2006), *Top. Appl. Phys*. **101**, 57-103 (2006).

9. Beach, G. S. D., Nistor, C., Knutson, C., Tsoi, M. & Erskine, J. L. Dynamics of field-driven domain-wall propagation in ferromagnetic nanowires. *Nature Mater*. **4**, 741-744 (2005).

10. Cowburn, R. & Petit, D. Turbulence ahead. *Nature Mater*. **4**, 721-722 (2005).

11. Nakatani, Y., Thiaville, A. & Miltat, J. Faster magnetic walls in rough wires. *Nature Mater.* **2**, 521-523 (2003).

12. Schryer, N. L. & Walker, L. R. The motion of 180° domain walls in uniform dc magnetic fields. *J. Appl. Phys*. **45**, 5406-5421 (1974).

13. Nakatani, Y., Thiaville, A. & Miltat, J. Head-to-head domain walls in soft nano-strips: a refined phase diagram. *J. Magn. Magn. Mater*. **290-291**, 750-753 (2005).

14. Kunz, A. Simulated domain wall dynamics in magnetic nanowires. *J. Appl. Phys*. **99**, 08G107 (2006).

15. Tomas, L. et al. Oscillatory dependence of current driven domain wall motion on current pulse length. *Nature,* **443**, 197-200 (2006).

16. Laufenberg, M. et al. Temperature dependence of spin-torque effect in current induced domain wall motion. *Phys. Rev. Lett*. **97**, 046602 (2006).

17. Hayashi, M. et al. Dependence of current and field driven depinning of domain walls on their structure and chirality in permalloy nanowires. *Phys. Rev. Lett*. **97**, 207205 (2006).

18. Hayashi, M., Tomas, L., Rettner, C., Moriya, R., & Parkin, S.S.P. Direct observation of the coherent precessionof magnetic domain walls propagating along permalloy nanowires. *Nature Physics*, **3**, 21-25 (2007).

19. See Fig. S3.

20. Thiele, A. A. Applications of the gyrocoupling vector and dissipation dyadic in the dynamics of magnetic domains. *J. Appl. Phys.* **45**, 377-393 (1974).

21. He, J., Li, Z. & Zhang, S. Current-driven vortex domain wall dynamics by micromagnetic simulations. *Phys. Rev. B* **73**, 184408 (2006).

22. Thiaville, A., Garcia, J. M. & Miltat, J. Domain wall dynamics in nanowires. *J. Magn. Magn. Mater.* **242-245**, 1061-1063 (2002).

23. Filippov, B. N., Korzunin, L. G. & Kassan-Ogly, F. A. Nonlinear dynamics of vortex-like domain walls in magnetic films with in-plane anisotropy. *Phys. Rev. B* **64**, 104412 (2001).

24. Chern, G.-W., Youk, H. & Tchernyshyov, O. Topological defects in flat nanomagnets: The magnetostatic limit. *J. Appl. Phys.* **99**, 08Q505 (2006).

25. Guslienko, K. Yu. *et al.* Eigenfrequencies of vortex state excitations in magnetic submicron-size disks. *J. Appl. Phys.* **91**, 8037-8039 (2002).

26. Guslienko, K. Yu., Han, X. F., Keavney, D. J., Divan, R. & Bader, S. D. Magnetic vortex core dynamics in cylindrical ferromagnetic dots. *Phys. Rev. Lett.* **96**, 067205 (2006).

27. Choe, S.-B. *et al.* Vortex core-driven magnetization dynamics. *Science* **304**, 420-422 (2004).

28. Tchernyshyov, O. & Chern, G.-W. Fractional vortices and composite domain walls in flat nanomagnets. *Phys. Rev. Lett.* **95**, 197204 (2005).

29. Tsoi, M., Fontana, R.E. & Parkin, S. S. P. Magnetic domain wall motion triggered by an electric current. *Appl. Phys. Lett.* **83**, 2617-2619 (2003).

30. Grollier, J. et al. Switching a spin valve back and forth by current-induced domain wall motion. *Appl. Phys. Lett.* **83.** 509-511 (2003).

31. Allwood, D. A. *et al.* Submicrometer ferromagnetic NOT gate and shift resister. *Science* **296**, 2003-2006 (2002).

32. Lee, J.-Y., Choi, S. & Kim, S.-K. Dynamics of transverse magnetic domain walls in rectangular-shape thin-film nanowires studies by micromagnetic simulations. *J. of Magnetics* **11**, 4-76 (2006).

33. Lee, K.-S., Choi, S. & Kim, S.-K. Radiation of spin waves from magnetic vortex cores by their dynamic motion and annihilation processes. *Appl. Phys. Lett.* **87**, 192502 (2005).

Acknowledgements

This work was supported by Creative Research Initiatives (Research Center for Spin Dynamics & Spin-Wave Devices) of MOST/KOSEF.

Competing financial interests

The authors declare that they have no competing financial interests.

Figure Legends

Figure 1. Magnetic-field-driven domain-wall motions. a, The rectangular-shaped nanostripes of thickness $h$=10 nm, length $l$=6 μm, variable widths $w$ as noted, and the coordinate system used. The local magnetization configurations at equilibrium for $w$=140 and 240 nm display a typical TW with the V-shaped polarization and a VW with the downward core orientation placed at the stripe center. An applied magnetic field, $H$, is indicated by the black arrow. The color wheel indicates the direction of the local magnetizations in the nanostripes. b, Average velocities of domain walls versus $H$ for two cases of $w$=140 nm (solid symbols) and $w$=240 nm (open symbols). The square, circle, and triangle symbols for each $w$ case, respectively, indicate Region I, the steady motion of a single domain wall, Region II, the oscillatory motion of different internal DW structures that are periodically transformed from one to another during the DW motion, and Region III, the magnetization reversal via multi-vortex (-antivortex) states. The insets show the representative snapshot images of the dynamic internal DW structures in the different characteristic regions for the case of $w$=140 nm, as example. c, Representative displacement-versus-time curves at the given fields of $H$= 5, 25, 30, 150 Oe in the different regions. The red straight lines are fitted to the data points within the range of 0 - 2 μm, indicating each average velocity $\bar{v}$ for the corresponding field. For more data in a wide range of $w$ and $H$, see Fig. S1.

Figure 2. Three unique periodicities of the dynamic transformations of the internal DW structures. a, The unique periods of the oscillatory D-vs-t curves. b, Energy variations with time during the periodic DW transformations. Each colour indicates the total $E_{tot}$, Zeeman $E_{zeem}$, exchange $E_{ex}$, and dipolar $E_{dip}$ energy terms, compared with the sum of $E_{zeem}$ and $E_{dip}$. The vertical red lines separate the individual regions of the DW transformations. c, Plane-view images of the local in-plane magnetization ($M_x / M_s$) components for the individual internal DW structures that transform from type one to another in a periodic manner, together with the trajectories of the TW linear motions, and the VW and AVW gyrotropic motions within the nanostripes. The trajectories were plotted by following the maximum exchange energy values at given times (marked in each region and each case indicated by the gray colors in a) for the three different cases of ($w$, $H$) = (140, 25), (240, 20), and (240 nm, 30 Oe). The solid circles and arrows on the TW types indicate the nucleation sites where the cores of VWs or AVWs are created, and the directions of their initial movements, respectively. The arrows on the trajectories indicate the directions of the movement of each DW type. The open gray-colored circles indicate the starting positions of the movements of each DW.

Figure 3. **Half-integer and integer topological charges of each type of DWs as magnetic solitions.** Topological charges of magnetic solitons within each DW are indicated by the symbols as noted. The sum of the topological charges of each DW is always zero. The solid-line triangles on the TW, VW, and AVW structures indicate the local magnetization configurations as in the TW type.

Figure 4. **Comparison of traveling times (time period) and trajectories between the simulations and analytical results.** a and b are the cases of *w*=140 and 240 nm, respectively. Upper row: the trajectories of the motions of the cores of VWs and AVWs obtained from the simulations (solid circles) are well fitted to the analytically derived parabolic orbits (green curves) described in the text. The fitting parameter of $\kappa$ is -1.8×10$^{-2}$ A$^2$ for AVW and +9.1×10$^{-3}$ A$^2$ for VW. Open square and triangle symbols in the bottom row indicate the simulated traveling times of the VW and AVW movements in the corresponding field. Those data points are fitted to $H^1$, which can be compared with the predicted traveling time, $t_w = \pi/\gamma H$ (orange-colored solid lines). Open circles indicate the TW traveling times between acts of the V (AV) emission/absorption.

Figure 1.

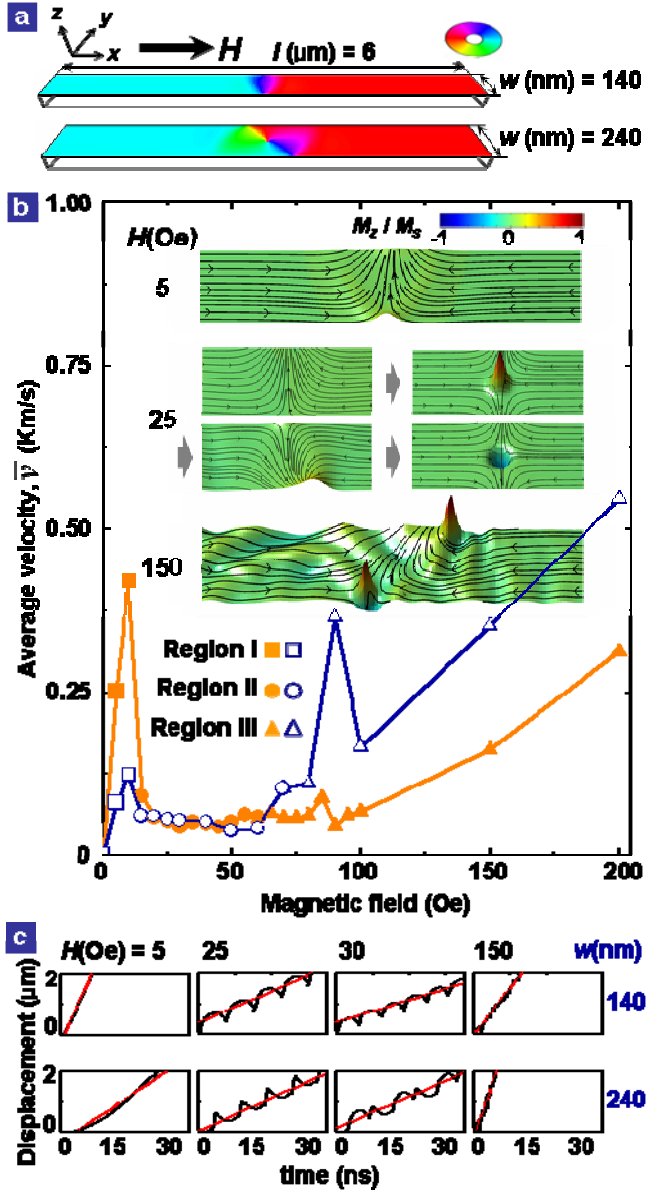

Figure 2.

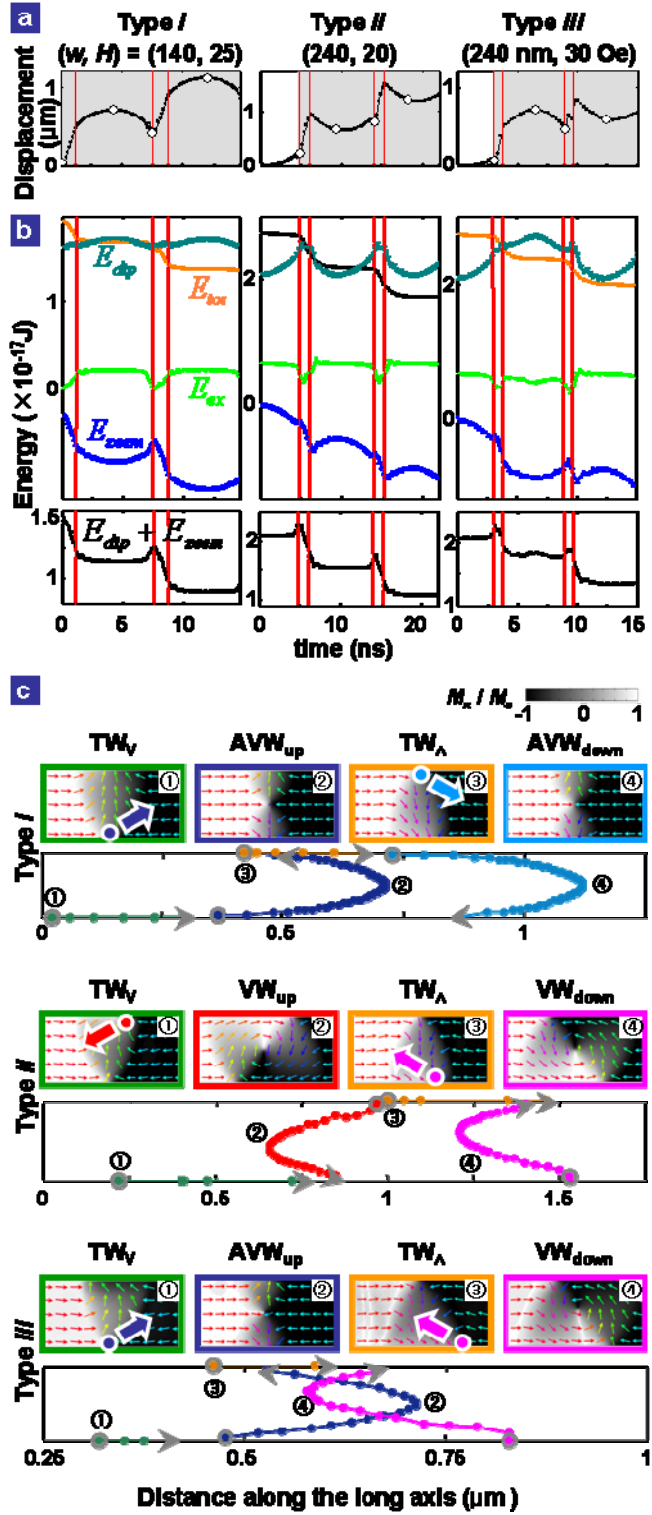

Figure 3.

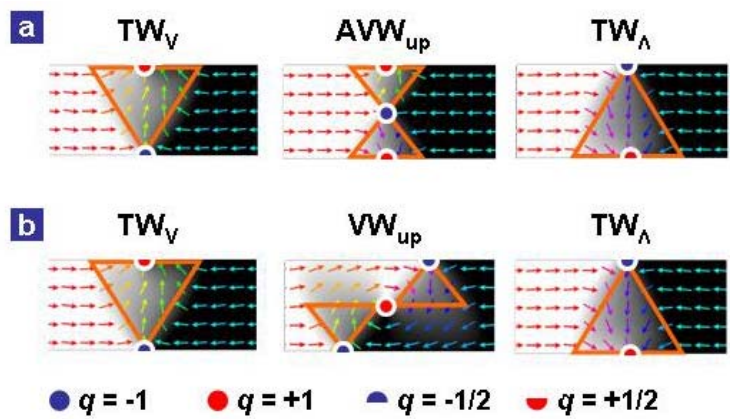

Figure 4

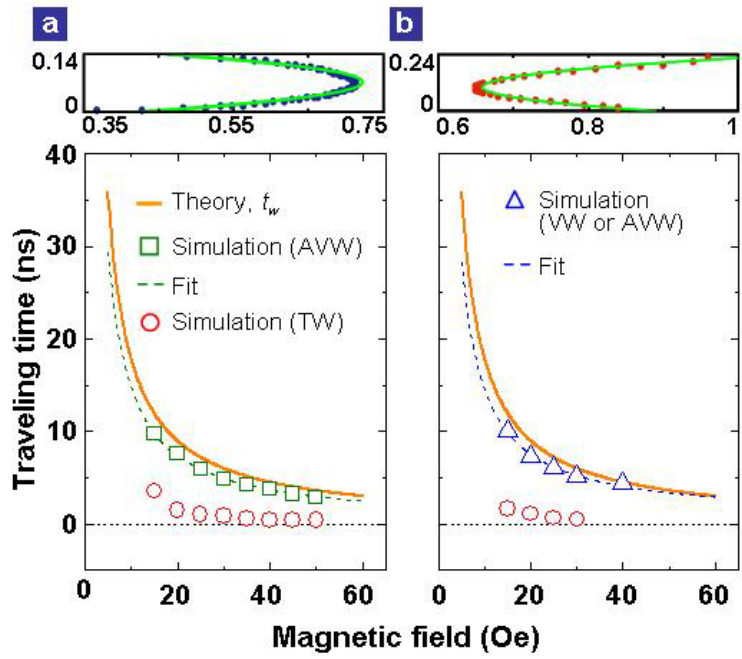